\documentclass[aps,prb,twocolumn,showpacs,superscriptaddress]{revtex4}
\usepackage{graphicx}
\begin{document}


\title{$\mu$SR studies of  
the frustrated quasi-2d square-lattice spin system 
Cu(Cl,Br)La(Nb,Ta)$_{2}$O$_{7}$:
evolution from spin-gap to antiferromagnetic state\/}
     \author{Y.~J.~Uemura}
     \altaffiliation[authors to whom correspondences should be addressed ]{}
     \affiliation{Department of Physics, Columbia University, New York, New York 10027, USA}
     \author{A.~A.~Aczel}
   \affiliation{Department of Physics and Astronomy, McMaster University, Hamilton, Ontario, 
Canada L8S 4M1}
     \author{Y.~Ajiro}
     \affiliation{Department of Chemistry, Kyoto University, Kyoto 606-8502, Japan}
     \author{J.~P.~Carlo}
     \affiliation{Department of Physics, Columbia University, New York, New York 10027, USA}
     \author{T.~Goko}
     \affiliation{Department of Physics, Columbia University, New York, New York 10027, USA}
     \affiliation{TRIUMF, Vancouver, British Columbia, Canada V6T 2A3}
     \author{D.~A.~Goldfeld}
     \affiliation{Department of Physics, Columbia University, New York, New York 10027, USA}
     \author{A.~Kitada}
    \affiliation{Department of Chemistry, Kyoto University, Kyoto 606-8502, Japan}
    \author{G.~M.~Luke}
    \author{G.~J.~MacDougall}
   \affiliation{Department of Physics and Astronomy, McMaster University, Hamilton, Ontario, 
Canada L8S 4M1}
       \author{I.~G.~Mihailescu}
    \affiliation{Department of Physics, Columbia University, New York, New York 10027, USA}
    \author{J.~A.~Rodriguez}
   \affiliation{Department of Physics and Astronomy, McMaster University, Hamilton, Ontario, Canada L8S 4M1}
       \author{P.~L.~Russo}
     \affiliation{Department of Physics, Columbia University, New York, New York 10027, USA}
    \author{Y.~Tsujimoto}
    \affiliation{Department of Chemistry, Kyoto University, Kyoto 606-8502, Japan}
    \author{C.~R.~Wiebe}
    \affiliation{Department of Physics, Florida State University, Tallahassee, Florida 32310, USA}
    \author{T.~J.~Williams}
   \affiliation{Department of Physics and Astronomy, McMaster University, Hamilton, 
Ontario, Canada L8S 4M1}
     \author{T.~Yamamoto}
     \author{K.~Yoshimura}
    \affiliation{Department of Chemistry, Kyoto University, Kyoto 606-8502, Japan}
    \author{H.~Kageyama}
     \altaffiliation[authors to whom correspondences should be addressed ]{}
    \affiliation{Department of Chemistry, Kyoto University, Kyoto 606-8502, Japan}
\date{\today}
\begin{abstract}
{We report muon spin relaxation ($\mu$SR) and magnetic susceptibility measurements on 
Cu(Cl,Br)La(Nb,Ta)$_{2}$O$_{7}$, which demonstrate:
(a) the absence of static magnetism in (CuCl)LaNb$_{2}$O$_{7}$ down to 15 mK confirming a
spin-gapped ground state;
(b) phase separation between partial volumes with a spin-gap and static magnetism in
(CuCl)La(Nb,Ta)$_{2}$O$_{7}$;
(c) history-dependent magnetization in the (Nb,Ta) and (Cl,Br)
substitution systems;  
(d) a uniform long-range collinear antiferromagnetic state in (CuBr)LaNb$_{2}$O$_{7}$; and
(e) a decrease of N\'eel temperature with decreasing Br concentration $x$ in 
Cu(Cl$_{1-x}$Br$_{x}$)LaNb$_{2}$O$_{7}$ with no change in the ordered Cu moment size
for $0.33 \leq x \leq 1$.  Together with several other $\mu$SR 
studies of quantum phase transitions in geometrically-frustrated 
spin systems, the present results reveal that the evolution 
from a spin-gap to a magnetically ordered state
is often associated with phase separation and/or a 
first order phase transition.\/}
\end{abstract}
\maketitle
\section{Introduction}

Modern studies of quantum phase transitions (QPTs) seek
novel features of ground states near phase boundaries. 
In a narrow range of $J_{2}/J_{1}$ ratios of square lattice spin systems,
geometrical frustration of the nearest neighbour ($J_{1}$) and next
nearest neighbour ($J_{2}$) exchange interactions is expected 
to yield a spin-gap state.  
Recent synthesis of a new square lattice system (CuCl)LaNb$_{2}$O$_{7}$ 
brought the first long-awaited system which might help elucidating 
this hypothesis and the associated QPT.  In this paper, we report 
muon spin relaxation ($\mu$SR) and low-field susceptibility studies
of this system and relevant compounds obtained by (Cl,Br) and (Nb,Ta) substitutions.

Unlike thermal phase transitions which represent changes of 
systems as a function of 
temperature, studies of QPTs follow the 
evolution of ground states across phase boundaries
at T $\rightarrow 0$ by varying, for example, chemical composition or 
pressure as a tuning parameter.
Recent experimental studies revealed novel and sometimes unexpected 
features of QPTs, 
such as first-order-like behavior, phase separation and slow spin fluctuations 
at boundaries of magnetically-ordered and paramagnetic states in itinerant electron magnets 
MnSi \cite{pfleidererMnSiPRB,pfleidererMnSiNature,uemuraNaturePhys} 
and (Sr,Ca)RuO$_{3}$ \cite{uemuraNaturePhys}, which are 
metallic systems without spin frustration.
Phase separation has also been found between the
stripe spin-charge ordered state and the
superconducting states without static magnetism in high-T$_{c}$ cuprate 
systems \cite{saviciLCOPRB,kojimaLESCOPhysica,mohottalaLSCONaturePhys}
involving holes doped in antiferromagnetic CuO$_{2}$ planes.
It is interesting to compare these results to QPTs of
insulating and geometrically frustrated spin 
systems (GFSS) \cite{ramirezreview,lhuillierreview} at boundaries between 
magnetically ordered states and spin-gap/spin-liquid states 
realized without static magnetism. 

GFSS on triangular or Kagom\'e lattices exclusively involve antiferromagnetic
nearest-neighbour interactions in hexagonal geometry.  In contrast, 
the two-dimensional (2-d) square-lattice $J_{1}$-$J_{2}$ system is 
unique in its underlying square lattice geometry as well as 
the possible involvement of ferromagnetic interactions.
Several compounds of vanadium oxide \cite{melziprl,geibelv}, 
synthesized in the early search for  
spin-gapped $J_{1}$-$J_{2}$ materials, unfortunately showed magnetic order at 
low temperatures (2.1-3.5 K).  As illustrated in Fig. 1c, 
the $J_{2}/J_{1}$ ratios of these systems,
estimated from magnetic susceptibility, narrowly missed the region predicted 
for formation of a singlet ground state 
with a spin-gap \cite{shannonepjb,shannonprb}.

\begin{figure}[t]
\includegraphics[width=3.4in,angle=0]{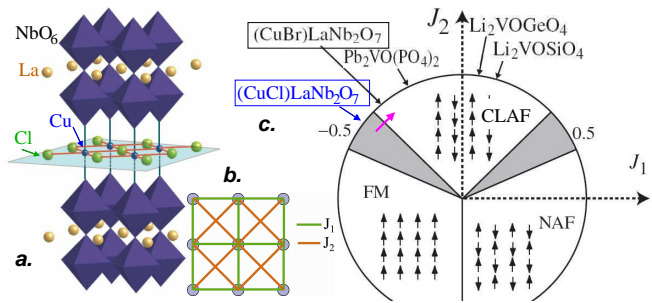}%
\label{Figure 1.} 
\caption{\label{Figure 1.}
(color)
(a) Crystal structure of (CuCl)LaNb$_{2}$O$_{7}$ \cite{kageyamajpsj2005}.
(b) Exchange interactions $J_{1}$ and $J_{2}$ on the 2-d square lattice.
(c) Conceptual phase diagram of the spin 1/2 square lattice $J_{1}$-$J_{2}$ 
model \cite{shannonprb,shannonepjb} as a function of $J_{1}$ and $J_{2}$
with regions of collinear antiferromagnetic (CLAF), ferromagnetic (FM) and 
N\'eel antiferromagnetic (NAF) order.  A spin-gap or spin-liquid state is expected in the shaded
region. The present work elucidates the evolution illustrated by the purple arrow.} 
\end{figure}

In 2005, Kageyama and co-workers \cite{kageyamajpsj2005} succeeded in 
synthesizing a spin-gap candidate $J_{1}$-$J_{2}$ system 
(CuCl)LaNb$_{2}$O$_{7}$, which has
a quasi 2-d crystal structure with S=1/2 Cu moments as shown in Figs. 1a and b.  
The magnetic susceptibility $\chi$ of this system, shown in Fig. 2, exhibits typical  
spin-gap behavior with an estimated gap $\Delta/k_{B} = 27$ K.  The spin gap was directly 
observed by inelastic neutron scattering \cite{kageyamajpsj2005} 
and high-field magnetization \cite{hfmagnetization}.  The very small
Weiss temperature $\Theta = J_{1}+J_{2} \leq 5$ K in the $1/\chi$ plot (Fig. 2)
indicates that $J_{1}$ and $J_{2}$ have nearly equal magnitudes but opposite signs.
Based on these observations, we assign this compound to fall in the spin-gap region in Fig. 1c. 

\begin{figure}[t]
\includegraphics[width=3.0in,angle=0]{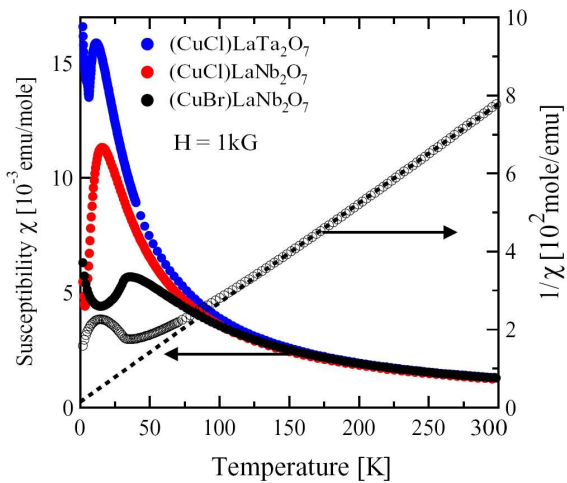}%
\label{Figure 2.} 
\caption{\label{Figure 2.}
(color)
Magnetic susceptibility $\chi$ of (CuCl)LaNb$_{2}$O$_{7}$, (CuCl)LaTa$_{2}$O$_{7}$ 
and $\chi$ and $1/\chi$ of (CuBr)LaNb$_{2}$O$_{7}$ in an external field of 1 kG.} 
\end{figure}

A sister compound (CuBr)LaNb$_{2}$O$_{7}$ orders
into a collinear antiferromagnet (CLAF) below $T_{N}$ = 32 K, as shown by 
neutron scattering \cite{brneutron} and susceptibility measurements (Fig. 2). 
This indicates a slight change of the $J_{2}/J_{1}$ ratio of the Br compound from that of
the Cl compound (Fig. 1c), together with a relatively large magnitude of
the antiferromagnetic $J_{2}$ which is at least enough to support this N\'eel
temperature.  In order to study the evolution from the CLAF to 
the spin-gap states, illustrated by a purple arrow in Fig. 1c, we
investigate solid solution systems Cu(Cl$_{1-x}$Br$_{x}$)LaNb$_{2}$O$_{7}$
as well as (CuCl)La(Nb$_{1-y}$Ta$_{y}$)$_{2}$O$_{7}$ \cite{kageyamaptps}.
The latter substitution is of great interest since it tends to suppress the spin gap, as seen
in $\chi$ for $y$ = 1 in Fig. 2, without perturbing the direct exchange path on the CuCl square
lattice plane. This may allow finer tuning of the $J_{2}/J_{1}$ ratios with 
smaller effect of randomness.  

We started a project to study these systems by $\mu$SR in 2005.  
Since then a few parallel
efforts have been made in neutron scattering, high-field susceptibility / magnetization,
NMR, and Raman scattering measurements.  While NMR and Raman studies have been / will be reported in 
separate independent papers \cite{nmrtakigawa,ramantbp} 
$\mu$SR and the remaining studies and characterization of the specimens
will be reported in three consecutive papers (I, II \cite{kitadanbta} and III 
\cite{tsujimotoclbr}).
In this paper (I), we focus on $\mu$SR and low-field magnetic susceptibility measurements on 
Cu(Cl,Br)La(Nb,Ta)$_{2}$O$_{7}$, which demonstrate:
(a) absence of static magnetism in (CuCl)LaNb$_{2}$O$_{7}$ down to 15 mK confirming the
spin-gapped ground state;
(b) phase separation between partial volumes with a spin-gap and static magnetism in
(CuCl)La(Nb,Ta)$_{2}$O$_{7}$;
(c) history-dependent magnetization in the (Nb,Ta) and (Cl,Br)
substitution systems; and 
(d) a uniform long-range collinear antiferromagnetic state in (CuBr)LaNb$_{2}$O$_{7}$.
Details of preparation of specimens,
elastic neutron scattering and high-field magnetization/susceptibility studies will be
reported in paper II for (Nb,Ta) substitutions and III for (Cl,Br) substitutions.
The companion paper II is submitted simultaneously with the present paper so that it may be
published back-to-back with the present paper.  Paper III will be published separately.

The development of theories for $J_{1}$-$J_{2}$ systems are in progress.  Following initial
conjecture \cite{shannonprb,shannonepjb} of existence of a spin gap in the border region of CLAF state and 
ferromagnetic (FM) state, a more recent theory \cite{shannonprl} predicts 
nematic spin arrangement, instead of the spin gap, at this border, 
when first and second neighbour interactions are considered in a Heisenberg model of 
a 2-dimensional square lattice.  However, a spin-gap state could well be expected 
for situations involving higher order interactions, such as, 3rd and 4th nearest neighbour exchange
interactions, anisotropy or three-dimensional interlayer couplings.   
In a recent experimental effort to develop compounds relevant to the present
Cu(Cl,Br)La(Nb,Ta)$_{2}$O$_{7}$ system, Tsujimoto et al. \cite{tsujimotoprb} synthesized  
(CuBr)A$_{2}$B$_{3}$O$_{10}$ (A = Ca, Sr, Ba, Pb; B = Nb, Ta) which has three perovskite 
layers between the adjacent CuBr planes.  One of these
systems, (CuBr)Sr$_{2}$Nb$_{3}$O$_{10}$, exhibits a positive Curie temperature when 
the high-temperature susceptibility $\chi$
is extrapolated to low temperatures in a plot of $1/\chi$ versus T.  This suggests that a 
fine tuning of parameters indeed brings the square lattice Cu(Cl,Br) plane into the side of FM
correlations across the spin gap region, as illustrated in Fig. 1c.  

In an NMR study of CuClLaNb$_{2}$O$_{7}$ Yoshida {\it et al.\/} \cite{nmrtakigawa} 
found a signature
of dimerization, although the corresponding superlattice satellite peaks of 
X-ray scattering are rather weak in intensity, and the assigned symmetry is not fully 
consistent with the results of recent Raman measurements \cite{ramantbp}.
In view of these developments, here we adopt
the $J_{1}$-$J_{2}$ model as an appropriate
starting point for discussions of Cu(Cl,Br)La(Nb,Ta)$_{2}$O$_{7}$, although there
may be some influence of higher order interactions, dimerization, and other effects existing
in real materials.    
Even when a small dimerization is essential for the formation of the spin gap,
all the experimental results described in this paper can still be regarded as elucidating
an interesting boundary between a spin-gap and CLAF state.

During the course of this study, we found a history dependence of the magnetic
susceptibility of Cu(Cl,Br)La(Nb,Ta)$_{2}$O$_{7}$ which sets in at $T \sim 7$ K in a low
applied field of $\sim$ 100 G.  This feature will be reported in section IV, following
the $\mu$SR results in section III.
A renewed interest in $J_{1}$-$J_{2}$ systems was generated by the recent 
discovery of La(F,O)FeAs superconductors
\cite{hosono} which have Fe moments in this geometry \cite{neutron,theory}.
In section V, we will compare the present results with $\mu$SR studies of 
other GFSS, including Kagome lattice systems, cuprate and other systems on body-centered tetragonal lattices, 
and FeAs superconductors.  

\section{Experimental methods}

Polycrystal specimens of 
solid solution systems Cu(Cl$_{1-x}$Br$_{x}$)LaNb$_{2}$O$_{7}$
and (CuCl)La(Nb$_{1-y}$Ta$_{y}$)$_{2}$O$_{7}$ \cite{kageyamaptps} 
were synthesized at Kyoto University using ion-exchange 
reactions
at low temperatures, as described in refs. \cite{tsujimotoclbr,kitadanbta}.
X-ray diffraction results show no trace of impurity phases for
the entire concentration regions $0 \leq x \leq 1$ and 
$0 \leq y \leq 1$ within the experimental detection limit.
In the case of (Cl,Br) substitutions, the homogeneous and random distribution of 
substituted atoms has been verified by the variation of the a- and c-axis
lattice constants against composition $x$ \cite{tsujimotoclbr} following Vegard's law.
For the (Nb,Ta) substitutions, 
similar information 
could not be obtained from X-ray diffraction due to nearly equal atomic radius of Nb and Ta.
Chemical homogeneity of both the (Br,Cl) and (Nb,Ta) samples have been checked using
energy dispersive spectroscopy (EDS) of transmission electron microscope (TEM)
measurements, which confirmed uniform solutions with the spatial resolution of 
10 nm \cite{tsujimotoclbr,kitadanbta}. 
The specimens were pressed into disc-shaped pellets
having typical dimensions of 10 mm in diameter and 1-2 mm in thickness.
The magnetic susceptibility of a small piece of each of these specimens was 
measured using a standard
Quantum Design SQUID magnetometer at Kyoto University.

$\mu$SR measurements were performed at TRIUMF, 
the Canadian National Accelerator Laboratory in Vancouver.
Polarized positive muons were
implanted one-by-one into the specimens mounted in 
a gas-flow He cryostat for measurements above $T$ = 2.0 K at the M20 or M15 channels, 
and in a dilution-refrgerator cryostat for those between 15 mK and 10 K at the M15 channel.
The time evolution $A(t)$ of the muon spin polarization was obtained from the 
time histograms $F(t)$ and $B(t)$ of the forward and backward counters as 
$$A(t) = A_{o}G(t) = [F(t) - B(t)]/[F(t) + B(t)],  \eqno{(1)}$$
where $G(t)$ represents the relaxation function defined with $G(0)$ = 1.
Details of the $\mu$SR methods can be found in refs. \cite{uemuraNaturePhys,saviciLCOPRB}.

\section{Experimental results: $\mu$SR}

\subsection{Zero-field $\mu$SR spectra: ground state}

Due to its superb sensitivity to static magnetic order in
spin systems with random / dilute / very small ordered moments,
$\mu$SR provides a very stringent test to verify the absence of 
active magnetism expected in spin-gap candidate systems \cite{mendelskagome,condmatlsco}.
Figure 3a shows the Zero-Field (ZF) $\mu$SR time spectra obtained for 
(CuCl)LaNb$_{2}$O$_{7}$ which exhibit a very slow relaxation 
at $T$ = 2 K and 15 mK.  This depolarization can be decoupled
by a small longitudinal field (LF) of 50 G.  
These features are expected for 
relaxation caused by static nuclear dipolar fields.
To ensure good heat conduction at 15 mK, we reproduced the results 
with another ceramic specimen containing 30\%\ Au powder by weight.
Thus, we confirmed the absence of static magnetic order in (Cu,Cl)LaNb$_{2}$O$_{7}$
down to 15 mK.  Statistical / systematic error of the measurement
gives an upper limit of less than 2 \%\ of 
the total volume $V_{M}$ with static magnetism.  

\begin{figure}[t]
\includegraphics[width=2.5in,angle=0]{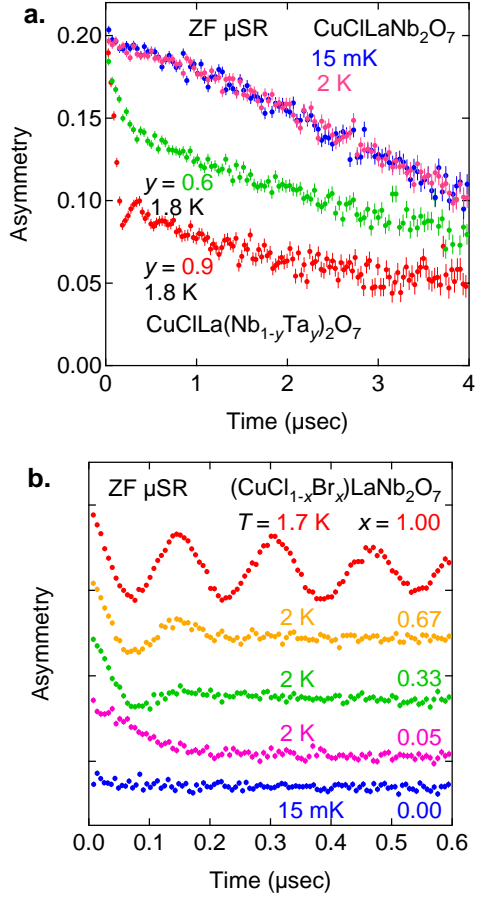}%
\label{Figure 3.} 
\caption{\label{Figure 3.} (color-online)
(a) Zero-field (ZF) $\mu$SR time spectra
in (CuCl)LaNb$_{2}$O$_{7}$ demonstrating absence of 
static magnetism, and in (CuCl)La(Nb$_{1-y}$Ta$_{y}$)$_{2}$O$_{7}$ showing magnetic order in
a partial volume fraction. (b) ZF-$\mu$SR spectra in (CuCl$_{1-x}$Br$_{x}$)LaNb$_{2}$O$_{7}$
at low temperatures.}
\end{figure} 

In contrast, fast decay of the ZF-$\mu$SR spectra was observed at $T$ = 1.8 K
in (Nb$_{1-y}$Ta$_{y}$) substitution systems with $y \geq $ 0.3.
With increasing $y$, the amplitude of the fast-relaxing component increases (Fig. 3a),
indicating the existence of static magnetism with increasing volume fraction $V_{M}$.  
In the CLAF system (CuBr)LaNb$_{2}$O$_{7}$,
ZF-$\mu$SR spectra $A(t)$ exhibit long-lived precession below $T_{N}$ (Fig. 3b)
which indicates the existence of a well-defined local field at the muon site expected for 
homogenous long-range order.
With decreasing Br composition $x$ in the (Cl$_{1-x}$Br$_{x}$) substitution,  
the internal field at $T \sim 2$ K 
becomes increasingly inhomogeneous as shown by the damping of the oscillation in Fig. 3b.  
For $x$ = 0.05, the inhomogeneous static local field
results in a ZF-$\mu$SR lineshape often seen in dilute alloy spin-glass 
systems \cite{uemuraspinglass}.  The change
in the initial damping rate in Fig. 3b between $x$ = 0.33 and 0.05 indicates 
that the spin structure / orientation
and/or ordered moment size changes between these two concentrations.
We confirmed a static origin of the observed fast relaxation in $x$ = 0.05 via 
decoupling in LF at $T$ = 15 mK. 

\subsection{Zero-field $\mu$SR spectra: temperature dependence}

Figure 4 shows the time spectra of ZF-$\mu$SR observed in 
(CuBr)LaNb$_{2}$O$_{7}$ and (CuCl)LaTa$_{2}$O$_{7}$ at several different
temperatures.  Coherent oscillations are observed in the 
signal below the N\'eel temperature
T$_{N}$ =  32 K for the former and 7 K for the latter system.  The spectra 
of (CuBr)LaNb$_{2}$O$_{7}$ fit well to 
$$G_{z}(t) = A_{1}[\cos(\omega t)]\exp(-\Lambda_{1}t)]$$
$$+ A_{2}[\exp(-\Lambda_{2}t)] + A_{3}[\exp(-\Lambda_{3}t]\eqno{(2)}$$ 
where the first term represents the oscillating component, the second term represents
a component showing a fast damping with $\Lambda_{2} \sim \omega$ and the third
term represents 
a persisting slowly relaxing signal with $\Lambda_{3} << \omega$.
The spectra of (CuCl)LaTa$_{2}$O$_{7}$ exhibit faster depolarization of the oscillation
and fit well to a Bessel function term plus a nearly constant term, 
$$ G_{z}(t) = A_{1}[(1/\omega t)\sin(\omega t)\exp(-\Lambda_{1}t)]$$ 
$$+ A_{3}[\exp(-\Lambda_{3}t)]. \eqno{(3)}$$
Previously, the Bessel function line shape was found in ZF-$\mu$SR spectra from systems 
having incommensurate spin-density-wave \cite{lpletmtsfprb} or stripe spin modulation 
\cite{saviciLCOPRB}.  Neutron scattering measurements of
(CuCl)LaTa$_{2}$O$_{7}$, however, found a commensurate CLAF state \cite{kitadanbta}.  
Thus the present results may simply be
due to highly-disordered short-range AF correlations.  

\begin{figure}[t]
\includegraphics[width=3.4in,angle=0]{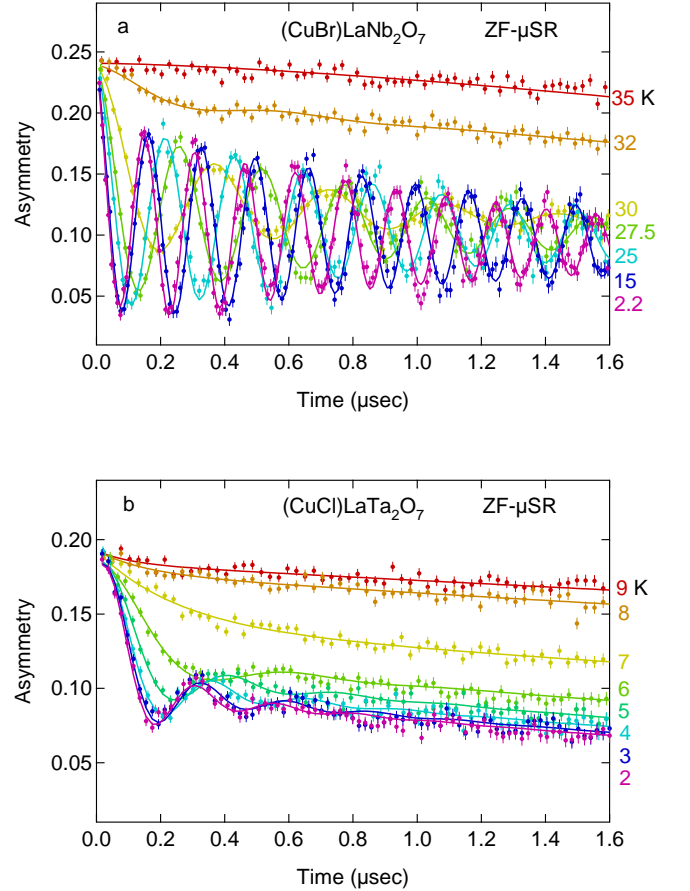}%
\label{Figure 4.} 
\caption{\label{Figure 4.} (color-online)
Zero-field (ZF) $\mu$SR time spectra
in (a) (CuBr)LaNb$_{2}$O$_{7}$ 
and (b) (CuCl)LaTa$_{2}$O$_{7}$
showing the onset of long-range antiferromagnetic order.
The solid lines represent eq. 2 in (a) and eq. 3 in (b).}
\end{figure}

In Fig. 5, we show the observed frequency $\nu = \omega/2\pi$ for the (Cl,Br) system with
$x$ = 1.0, 0.67, 0.33 and (CuCl)LaTa$_{2}$O$_{7}$.  The spectra from the latter
three systems were fit to Bessel functions, in view of significantly better fits as compared to 
the cosine function, although none of these systems exhibit a clear
indication of incommensurate spin correlations in neutron scattering.
With decreasing Br composition $x$, $T_{N}$ decreases, but the frequency $\nu(T\rightarrow 0)$
remains nearly unchanged at $0.33 \leq x \leq 1$.  
This indicates that the size of the ordered Cu moment does
not depend on $x$.  The frequency $\nu(T\rightarrow 0)$ for (CuCl)LaTa$_{2}$O$_{7}$
is significantly different from that of the other systems in Fig. 5, despite the fact that the same ordered
Cu moment size $\sim 0.6 \mu_{B}$ was reported by 
neutron scattering studies \cite{kitadanbta,tsujimotoclbr}.  
The lower frequency / internal field was
also found in all the (Nb,Ta) systems with $0.3 \leq y \leq 1$ 
(as discussed later) as well as in the (Cl,Br) system
with $x$ = 0.05 (see Fig. 3b).  Interestingly, these systems with 
the lower internal field all have a 
N\'eel temperature of $T_{N} \sim 7$ K.  These observations suggest that compounds with
sufficiently reduced $T_{N}$ have a short-ranged spin structure and a possibly different direction of the ordered Cu moments 
compared to systems with higher $T_{N}$.

\begin{figure}[t]
\includegraphics[width=3.4in,angle=0]{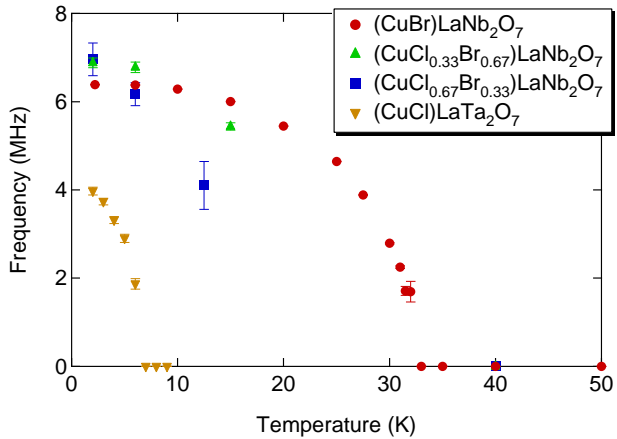}%
\label{Figure 5.} 
\caption{\label{Figure 5.} (color-online)
Muon spin precession frequency $\nu$ observed in 
Cu(Cl$_{1-x}$Br$_{x}$)LaNb$_{2}$O$_{7}$
with $x$ = 1.0, 0.67 and 0.33, and in 
(CuCl)LaTa$_{2}$O$_{7}$}
\end{figure} 

\subsection{Results in Weak Transverse Field: volume fraction of the magnetically ordered region}
   
To determine the volume fractions of regions with and without static magnetic order,
$\mu$SR measurements in weak transverse field (WTF) are quite useful, as shown in 
ref \cite{uemuraNaturePhys}. The persistent oscillation amplitude 
in WTF reflects muons landing in a paramagnetic 
or non-magnetic environment.  Figures 6a and b show the precessing amplitudes 
in (Nb$_{1-y}$Ta$_{y}$) and (Cl$_{1-x}$Br$_{x}$) systems mostly in WTF = 100 G.
Some of the data were obtained with WTF $\sim$ 50 and 30 G due to limitation of
available spectrometers.
A sharp onset of the CLAF ordered state is seen for the Br substitutions with 
$x$ = 0.2 - 1 (Fig. 6b) at temperatures consistent with the magnetic susceptibility results 
reported in paper III \cite{tsujimotoclbr}.
The (Nb,Ta) systems show static magnetism below a common
onset temperature $T_{N}\sim$ 6-7 K, with the paramagnetic volume fraction decreasing gradually
with decreasing temperature towards the partial fraction dependent on $y$.
This confirms the phase separation observed in ZF-$\mu$SR.
In both Figs. 6a and b, about 25\%\ of muons remain in a paramagnetic environment
even when static magnetism is established in the full volume
fraction as in the CLAF state of (CuBr)LaNb$_{2}$O$_{7}$ \cite{brneutron}.  
These muons are likely occupying sites where local fields from antiferromagnetic
Cu spins nearly cancel for symmetry reasons.  
Judging from the amplitude from the remaining 75\%\  of muons, we conclude that 
systems with $x \geq 0.2$ order in a full volume fraction $V_{M}$ = 1.0, while those with 
$x$ = 0.05 and $0.3 \leq y < 0.9$ undergo phase separation between ordered and spin-gapped
volumes.  The spin-gap volume prevails to nearly the full fraction for $y < 0.3$.

\onecolumngrid

\begin{figure}[h]

\begin{center}
\includegraphics[angle=0,width=6.0in]{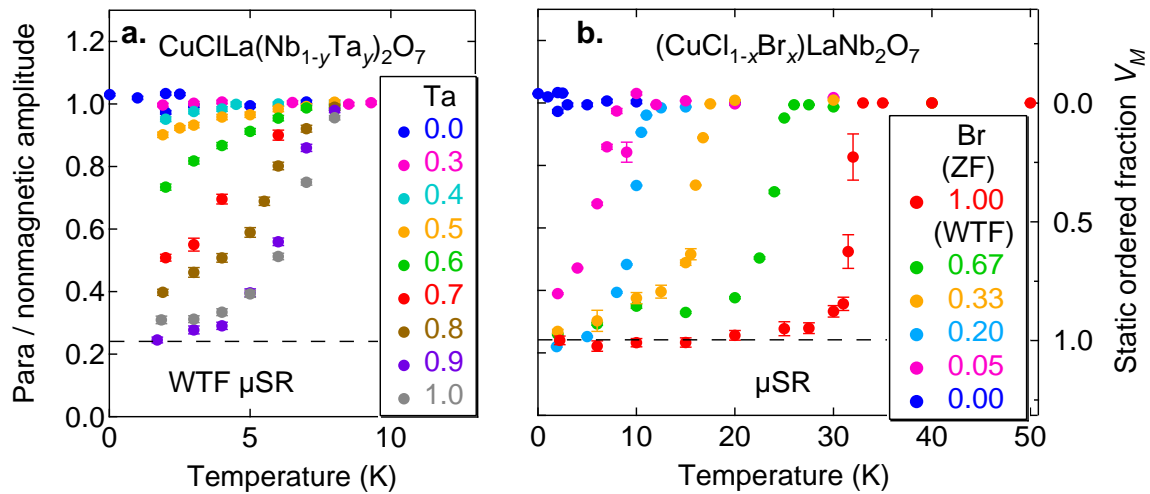}
\label{Figure 6.} 

\caption{\label{Figure 6.}
(color)
(a) and (b): Amplitude of persisting muon spin precession in a weak transverse field (WTF) of 
100 G in (a) (CuCl)La(Nb,Ta)$_{2}$O$_{7}$ and (b) Cu(Cl,Br)LaNb$_{2}$O$_{7}$.  This represents
the fraction of  muons landing in para- or non-magnetic environment.  In each specimen, about 25\%\ of the amplitude
persists at any temperature and composition as denoted by the 
broken line, which
presumably comes from muons at sites where 
the local fields from Cu moments cancel via symmetry reasons.}
\end{center}
\end{figure}

\twocolumngrid

\subsection{ZF-$\mu$SR spectra in phase-separated systems}

\onecolumngrid

\begin{figure}[t]
\begin{center}
\includegraphics[angle=0,width=6.0in]{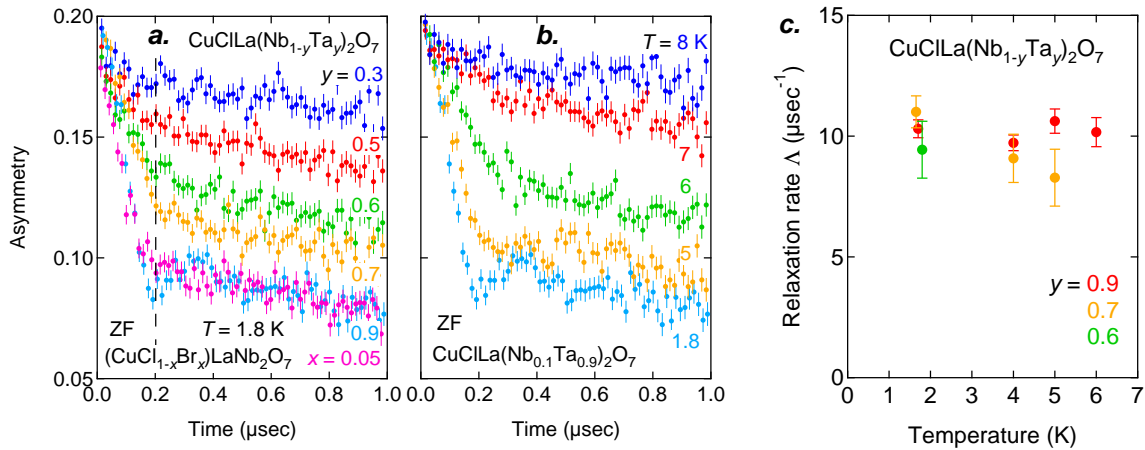}
\label{Figure 7.} 
\caption{\label{Figure 7.}
(color)  System (a) and temperature (b) dependence of the $\mu$SR time spectra in zero field
in (CuCl)La(Nb,Ta)$_{2}$O$_{7}$, which exhibit signals of fast- and slow-decay components,
with composition- and $T$-dependent amplitudes.  (c) The exponential muon spin 
relaxation rate $\Lambda$ of the 
fast decay component in (a) and (b).}
\end{center}
\end{figure}
\twocolumngrid
Static magnetism in the phase-separated region can be elucidated by 
ZF-$\mu$SR spectra in Figs. 7a and b for quantum (Fig. 7a) and thermal (Fig. 7b) evolution in
the (Nb,Ta) systems.  The similarities of these two figures demonstrates that
near the phase boundary between CLAF and spin-gapped states, both quantum and 
thermal transitions involve phase separation with gradual change of $V_{M}$
as a function of  temperature $T$ and composition $y$.   
These spectra can be decomposed into two components with 
fast and slow relaxation, respectively, having $T$- and $y$-dependent
amplitude ratios.  
The exponential relaxation rate $\Lambda$ of the fast component is nearly independent of $T$ and $y$,
as shown in Fig. 7c, 
which implies that the ordered regions in different $T$ and $y$ share common
microscopic spin arrangements of Cu moments.  
The common decay rate, indicative of the same spin configuration, is also found in the
$x$ = 0.05 (Br,Cl) substitution system (Fig. 2c).
The present experiment, however, does not allow us to estimate the typical size
of the ordered regions.  

\section{Experimental results: low-field magnetic susceptibility}

Motivated by the ZF-$\mu$SR line shapes in Fig. 7 which resemble those expected 
in spin-glass systems \cite{uemuraspinglass}, we performed measurements of dc-magnetization $M$
in a weak field of 100 G with both Field Cooling (FC) and 
Zero-Field Cooling (ZFC).  As shown in Fig. 8a, marked departure of 
$M_{FC}$ from $M_{ZFC}$ sets in at $T$ = 6 K, coinciding with 
the onset of static magnetism 
in the (Nb$_{1-y}$Ta$_{y}$) systems.  
History dependence of $M$ was also 
found in (Cl$_{1-x}$Br$_{x}$) systems with $x \leq$ 0.66, with a common onset
temperature of 6 K, well below $T_{N}$ for $x$ = 0.66, 0.33 and 0.2 (Fig. 8b).  
The magnitudes of (M$_{FC}$-M$_{ZFC}$) at $T$ = 2 K for all of these systems
correspond to a very small ferromagnetic polarization $\leq 10^{-3} \mu_{B}$ 
per Cu (insets of Fig. 8a and b), nearly independent of the field for the 
FC measurements in 100 - 800 G in the $x$ = 0.05 system (inset of Fig. 8b).  

\begin{figure}[thb]
\includegraphics[width=3.0in,angle=0]{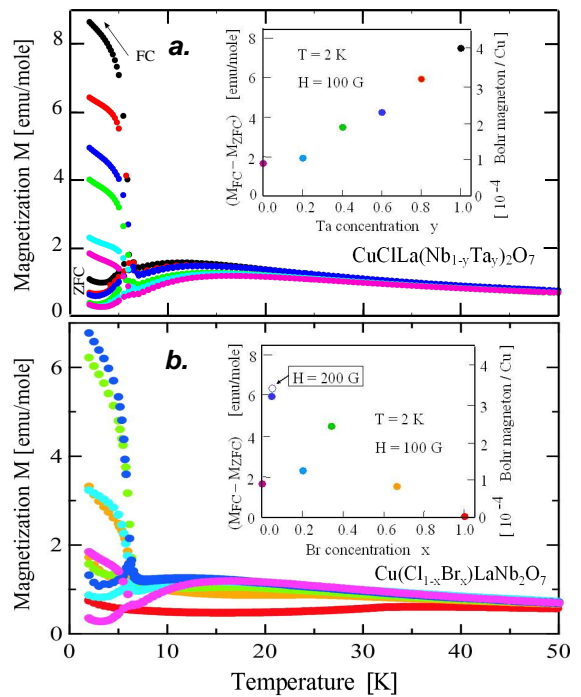}%
\caption{\label{Figure 8.}
(color)
(a) and (b): Magnetic susceptibility for both Field Cooling (FC) and Zero-Field Cooling (ZFC) in 
100 G obtained in (a) (CuCl)La(Nb,Ta)$_{2}$O$_{7}$ and (b)Cu(Cl,Br)LaNb$_{2}$O$_{7}$.
The inset figures show the irreversible magnetization $M_{irr} \equiv M_{FC}-M_{ZFC}$
at $T$ = 2 K.}
\end{figure}

The irreversible magnetization $M_{irr} \equiv (M_{FC}-M_{ZFC})$ in 
(CuCl)La(Nb$_{1-y}$Ta$_{y}$)$_{2}$O$_{7}$
at $T$ = 2 K (Fig. 8a) roughly scales with the volume fraction $V_{M}$ 
of the magnetically-ordered region (Fig. 6a). $M_{irr}$ and $V_{M}$ share the same 
onset temperature.  
These features suggest that the observed signal likely comes from the 
bulk volume of the region with static magnetism, rather than from dilute
ferromagnetic impurities.
The very small net ferromagnetic polarization (inset in Fig. 8a), together with 
the static Cu moment size of $\sim$ 0.6 $\mu_{B}$, indicate
mostly antiferromagnetic or random spin configurations,
with a very small ferrimagnetic / canted component.
Without distinguishing between these, we term the static
magnetism in the (Nb,Ta) system as a ``glassy antiferromagnetic'' (GAF) state.  
The relatively large $M_{irr}$ observed in the $y$=1 pure
Ta material rules out the notion that the irreversibility requires randomness
and/or a non-stoichiometic solid solution.  
Small, yet somewhat surprising non-zero values of $M_{irr}$ in the  
$y$ = 0 and 0.2 systems may be due to increasing ferrimagnetic / canted contributions, 
possibly from remaining static regions with a volume fraction $\leq$ 2\%.

\section{Phase diagram}

Summarizing these findings, we present a phase diagram in Fig. 9 as functions of 
the substitution concentrations $x$ and
$y$.  The phase-separated region with a partial volume fraction is
illustrated by a striped pattern, while regions with a full volume fraction are indicated by solie colors.
The first-order thermal transition is shown by the broken line,
while the solid line indicates the transitions which are likely second-order.  
The internal field observed in 
the CLAF state (pink) is significantly larger than 
that in the glassy antiferromangetic (GAF) state (green), suggesting that
these states are distinct.
Further characterization is required to clarify 
the coexistence of these two states in the region of $.05 \leq x \leq 1$.
The purple and orange arrows indicate, respectively, the history-dependent (HD) region 
where $M_{FC} \neq M_{ZFC}$ and the region with phase-separated static magnetism 
in a partial volume fraction (PV). 

\begin{figure}[tbh]
\includegraphics[width=3.3in,angle=0]{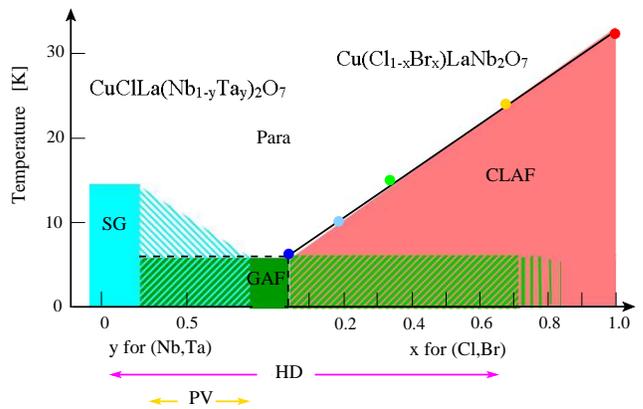}%
\caption{\label{Figure 9}(color)
Phase diagram of Cu(Cl,Br)La(Nb,Ta)$_{2}$O$_{7}$ obtained in the present study,
with spin-gap (SG), glassy antiferromagnetic (GAF) and collinear antiferromagnetic (CLAF) 
states. The striped region indicates phase separation.  The broken and solid 
lines denote, respectively, 
first- and second-order thermal transitions.  The arrows attached to the horizontal
axis show the region with history dependence (HD) of the magnetic susceptibility, and 
the region where static magnetism exists in a partial volume fraction (PV).}
\end{figure}

\section{Discussion}

The present $\mu$SR results have demonstrated that (CuCl)LaNb$_{2}$O$_{7}$ indeed posseses a non-magnetic
ground state, consistent with spin-gap formation, over the full volume fraction.
$\mu$SR has a superb sensitivity to static magnetism, as 
even small nuclear dipolar fields
can easily be detected.  Previous $\mu$SR measurements on some systems widely conceived to have 
spin-gap ground states, such as SrCu$_{2}$(BO$_{3}$)$_{2}$ \cite{C126}, CaV$_{2}$O$_{5}$ and
CaV$_{4}$O$_{9}$ \cite{C104}, resulted in the detection of muon spin relaxation persisting
to low temperatures.   
Contrary to these cases, the present study
established the absence of any detectable static and dynamic magnetism from Cu moments,
and demonstrated that the gapped state in (CuCl)LaNb$_{2}$O$_{7}$ is really robust.

Our results also revealed phase separation between volumes with and without static magnetism
in (CuCl)La(Nb$_{1-y}$Ta$_{y}$)$_{2}$O$_{7}$ for the Ta concentration $y > 0.3$.
In the companion paper II, Kitada et al. \cite{kitadanbta} report neutron scattering 
results of the (Nb,Ta) systems, where
the volume fraction was estimated from the  
intensities of the magnetic Bragg peaks.    
Kitada {\it et al.\/} also decomposed the
response of high-field magnetization into the ``gapped'' and ``ungapped'' signals, 
using the results of the
$y=0.4$ compound as the reference for the former and $y=1$ for the latter, and estimated 
the volume fraction of
the ungapped signal.  The volume fraction values derived from
the neutron and magnetization results in these procedures 
agree well with $V_{M}$ from $\mu$SR.  

Both neutron and magnetization studies detect
magnetism as a volume-integrated quantity, and they cannot distinguish a small volume
with a large individual moment versus a large volume with a small moment.  In contrast, 
$\mu$SR and NMR produce distinguishable signals from 
magnetically-ordered and paramagnetic regions, with the amplitudes proportional to 
corresponding volumes: the volume information is decoupled from 
that of the moment size.  Consequently, the real-space probes 
$\mu$SR and NMR have genuine advantages over neutron and magnetization in the
determination of ordered volume fractions.  In the present (Nb,Ta) systems, the 
size of the ordered moment does not depend on the Ta concentration $y$, 
as demonstrated in Fig. 7.
Due to this feature, the neutron and magnetization results for the volume fraction 
agreed well with the $\mu$SR results.

In paper III, Tsujimoto {\it et al.\/} \cite{tsujimotoclbr}
find signatures in the magnetic susceptibility
for the (Cl,Br) systems from which they derived the N\'eel temperature
quantitatively consistent
with the present $\mu$SR results.
Since the positive muon is a charged probe, there remains some suspicion that the
$\mu$SR results may be different from those of the bulk system due to possible
perturbation caused by the presence of the muon.  The good agreements of the 
volume fraction and $T_{N}$ with the neutron and susceptibility / magnetization 
results confirm that the muons are indeed probing bulk properties in the present systems.
The $\mu$SR results in Fig. 3(b) indicate increasingly short-ranged
spin correlations with decreasing Br concentration $x$.  This feature was not detected by
neutrons, due presumably to limited instrumental resolution and the weak signal expected from the 
magnetic Bragg reflections for powder specimens.  The spin correlation length 
remains to be determined by neutron studies in the future using single crystal specimens. 

\begin{figure}[t]
\includegraphics[width=2.5in,angle=90]{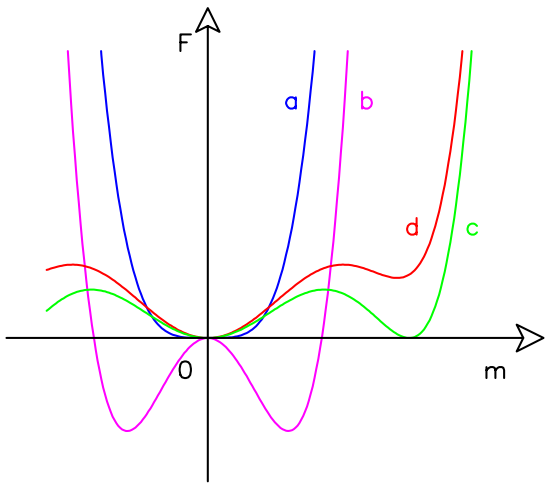}%
\caption{\label{Figure 10}(color)
A schematic view of free energy profile as a function of 
magnetic order parameter $m$.
The curves (a) and (b) represent the case of standard second-order phase transitions,
above and below $T_{c}$, respectively, while (d) and (c) for the first-order transitions
with (d) above $T_{c}$ and (c) at $T_{c}$.}
\end{figure}

History dependence (HD) of the magnetization 
$M$ has also been observed in the Kagom\'e lattice antiferromagnets
SrCr$_{x}$Ga$_{12-x}$O$_{19}$ (SCGO) \cite{ramirezscgo}, Cr-jarosite \cite{kerenjarosite} and 
many other GFSS, but 
without a jump in $M_{FC}$ at the onset temperature. 
The $M_{FC}$ jump observed in the present $J_{1}$-$J_{2}$
systems may be related to  
the involvement of the ferromagnetic exchange interaction $J_{1}$, which does not
exist in other GFSS based exclusively on antiferromagnetic couplings.   
The partial volume fraction (PV) of the magnetically-ordered region adjacent to the 
spin-gap region has also been observed recently by $\mu$SR studies in the 
Kagom\'e lattice Herbertsmithite system \cite{mendelskagome}.
Figure 10 shows a schematic view of free energy variation.
A free-energy profile with multiple minima at the order-parameter ($m$) values of 
zero and a finite value in Fig. 10 (lines c and d) can explain the origin of a first-order 
transition which often results in phase separation.  History dependence
can be caused by multiple free energy minima,
as generally seen in spin glasses with complicated free-energy landscapes
with many minima.
The present results show, however, that the regions for these two phenomena 
(purple and orange lines in Fig. 9) do not completely overlap.

When we look into previous $\mu$SR and susceptibility results in 
frustrated spin systems, we notice three patterns:
(a) Spin freezing at $T_{g}$, associated with critical slowing down (maxima of $1/T_{1}$
of $\mu$SR), history dependence of $M$ below T$_{g}$, and disappearance of dynamics
at $T\rightarrow 0$:  canonical spin glasses AuFe and CuMn 
\cite{uemuraspinglass}.
(b) Slowing down of spin fluctuations towards $T_{g}$, history dependence of $M$ 
below $T_{g}$, followed by persistent dynamic effects at $T\rightarrow 0$, $M$ remaining
finite at $T\rightarrow 0$:
Kagom\'e lattice systems SCGO SrCr$_{8}$Ga$_{4}$O$_{19}$ \cite{uemurascgo}, 
Cr-jarosite KCr$_{3}$(OH)$_{6}$(SO$_{4}$)$_{2}$ \cite{kerenjarosite}, 
and volborthite (Cu$_{x}$Zn$_{1-x}$)$_{3}$V$_{2}$O$_{7}$(OH$_{2}$) 2H$_{2}$O
\cite{fukayavolbor}, 
and a body-centered-tetragonal (BCT) system CePt$_{2}$Sn$_{2}$
\cite{cept2sn2}.  (c) Fully gapped state without static or dynamic magnetism at $T\rightarrow 0$
with clear reduction of $M$ at $T\rightarrow 0$ suggesting spin-gap
formation in the unperturbed system,
and appearance of phase-separated static magnetism when composition is varied:
CuClLaNb$_{2}$O$_{7}$ and the Kagom\'e system 
Herbertsmithite Zn$_{x}$Cu$_{4-x}$(OH)$_{6}$Cl$_{2}$ \cite{mendelskagome}.

We also note that many of the ``spin-gap candidate'' systems, such as 
SrCu$_{2}$(BO$_{3}$)$_{2}$ \cite{C126}, CaV$_{2}$O$_{5}$ and
CaV$_{4}$O$_{9}$ \cite{C104} and a doped Haldane gap system 
(Y,Ca)$_{2}$BaNiO$_{5}$ \cite{P58} exhibit responses of the pattern (b),
while the results of cuprate 
\cite{saviciLCOPRB,kojimaLESCOPhysica}
and FeAs superconductors \cite{gokocondmat,drewnatmaterial}
correspond to the pattern (c).
These observations indicate that near the fully spin-gapped 
(non-magnetic and/or superconducting) state,
systems have two choices: either to phase separate and exhibit magnetic 
order in a partial volume fraction (pattern (c)) or to become a strange
``spin-liquid'' which shows persistent dynamic spin responses at $T\rightarrow 0$
in the whole volume (pattern (b)).  Many of the $\mu$SR spectra observed in 
pattern (b) exhibit strange Gaussian line shapes which are very hard to 
decouple by longitudinal fields.  Further studies are required to clarify
which parameters are essential for a given system to follow pattern (b)
or (c), and to characterize more details of spin dynamics 
and line shapes for pattern (b).
  
Recent $\mu$SR results in MnSi and (Sr,Ca)RuO$_{3}$ \cite{uemuraNaturePhys} 
revealed phase separation at 
QPTs, very similar to the present results.  
Some history dependence was also noticed in MnSi \cite{pfleiderersans} near the phase boundary
where static magnetism disappears.  
Slow and presumably quantum spin fluctuations persisting at
very low temperatures have been observed 
in MnSi \cite{pfleidererMnSiNature,uemuraNaturePhys} near
the quantum phase boundary between the helically ordered and paramagnetic states.
Phase separation in HTSC cuprates was observed in
the QPT between spin-charge stripe and superconducting
states \cite{saviciLCOPRB,kojimaLESCOPhysica,mohottalaLSCONaturePhys}.  
The magnetic resonance mode in HTSC is a quantum slow
spin fluctuation and a soft-mode related to competing states across the quantum phase
boundary \cite{uemuraroton,uemurasces2008}.
Although a conclusive argument requires further accumulation of 
results on history dependence and inelastic soft modes in various systems, 
these observations hint that first order transitions, phase separation, 
history dependence, slow soft-mode spin fluctuations, and 
multiple free-energy minima may be common phenomena
generic to QPTs in systems both with and without geometrical frustration.  
Comprehensive theoretical studies on this aspect may
reveal further fascinating features.


We acknowledge financial support from
NSF DMR-05-02706 and DMR-08-06846
(Materials World Network, Inter-American Materials
Collaboration program), NSF DMR-01-02752 and DMR-02-13574 (MRSEC) at Columbia;
NSERC regular and CIAM supports and CIFAR (Canada) at McMaster; and 
the Japan-U.S. Cooperative Science
Program "Phase separation near quantum critical point in low-dimensional
spin systems" (Contract No. 14508500001) from JSPS of Japan and NSF, and 
Science Research on Priority Area (No. 19052004 and No.
16076210) from MEXT of Japan and GCOE program at Kyoto University.
We have greatly benefited from discussions with M.J.P. Gingras, A.J. Millis and N. Shannon.
\\

\vfill \eject

\end{document}